\documentclass[aps,twocolumn,prl,groupedaddress,showpacs]{revtex4-1}

\usepackage{graphicx,float}
\usepackage[normalem]{ulem}
\usepackage{color}
\usepackage[colorlinks=true,citecolor=blue,urlcolor=blue]{hyperref}

\usepackage{amsmath}
\usepackage{multirow}
\usepackage{booktabs}

\newcommand{\ket}[1]{|#1\rangle}
\newcommand{\boldvec}[1]{ \mathbf{#1} }

\begin{document}
\title{Correlated sensing with a solid-state quantum multi-sensor system for atomic-scale structural analysis}

\author{Wentao Ji$^{1,2,3}$$^{\S}$}
\author{Zhaoxin Liu$^{1,2}$$^{\S}$}
\author{Yuhang Guo$^{1,2}$$^{\S}$}
\author{Zhihao Hu$^{1,2}$}
\author{Jingyang Zhou$^{1,2}$}
\author{Siheng Dai$^{1,2}$}
\author{Yu Chen$^{1,2}$}
\author{Pei Yu $^{1,2}$}
\author{Mengqi Wang $^{1,2}$}
\author{Kangwei Xia$^{1,2}$}
\author{Fazhan Shi$^{1,2,3}$}
\author{Ya Wang$^{1,2,3}$}
\email{ywustc@ustc.edu.cn}
\author{Jiangfeng Du$^{1,2,3,4}$}
\email{djf@ustc.edu.cn}

\affiliation{$^1$ CAS Key Laboratory of Microscale Magnetic Resonance and School of Physical Sciences, University of Science and Technology of China, Hefei 230026, China}
\affiliation{$^2$ CAS Center for Excellence in Quantum Information and Quantum Physics, University of Science and Technology of China, Hefei 230026, China}
\affiliation{$^3$ Hefei National Laboratory, University of Science and Technology of China, Hefei 230088, China}
\affiliation{$^4$ School of Physics, Zhejiang University, Hangzhou 310027, China}
\affiliation{$^{\S}$ These authors contributed equally to this work}

\date{\today}
\begin{abstract}
	
\textbf{
Developing superior quantum sensing strategies ranging from ultra-high precision measurement to complex structural analysis is at the heart of quantum technologies.
While strategies using quantum resources, such as entanglement among sensors, to enhance the sensing precision have been abundantly demonstrated \cite{SQL_ion_2004, SQL_photon_2007, quantum_metrology_2011, SQL_atom_2016,SQL_optomechanical_2019}, the signal correlation among quantum sensors is rarely exploited \cite{covariance_2022}.
Here we develop a novel sensing paradigm exploiting the signal correlation among multiple quantum sensors to resolve overlapping signals from multiple targets that individual sensors can't resolve and complex structural construction struggles with.
With three nitrogen-vacancy centers as a quantum electrometer system, we demonstrate this multi-sensor paradigm by resolving individual defects' fluctuating electric fields from ensemble signals. 
We image the three-dimensional distribution of 16 dark electronic point-defects in diamond with accuracy approaching 1.7 nm via a GPS-like localization method.
Furthermore, we obtain the real-time charge dynamics of individual point defects and visualize how the dynamics induce the well-known optical spectral diffusion.
The multi-sensor paradigm extends the quantum sensing toolbox and offers new possibilities for structural analysis.
}

\end{abstract}

\maketitle

The advances in quantum sensing have revolutionized modern measurement technologies for various physical quantities \cite{SQL_ion_2004, SQL_photon_2007, quantum_metrology_2011, SQL_atom_2016, SQL_optomechanical_2019, SQUID_book_2003, SQUID_RSI_2006, atomic_vapour_2007, LIGO_2013, clock_rmp_2015, clock_network_2021, clock_gravity_2022, quantum_sensing_2017}.
Apart from deepening the understanding of fundamental physical laws and developing high-precision measurement, recent advancements of a single nanoscale quantum sensor have created the prospect of atomic-scale complex structural analysis.
An extraordinary example is the single nitrogen-vacancy (NV) center in diamond as a nanoscale quantum sensor \cite{nanoscale_nmr_2013, nanoscale_nmr2_2013, single_spin_esr_2015,2D_sensing_2017, 2D_nmr_2019, electrical_sensing, single_charge_2014, electrical_sensing_du, electrical_sensing_2008, electrical_sensing_2011, electrical_sensing_2018, electrical_sensing_2021} with high sensitivity of a single spin and fundamental charge \cite{single_spin_esr_2015, single_charge_2014, charge_transport_2021}, which is promising for atomic-scale magnetic structural analysis \cite{nanoNMR_2017, submHz_2017, high_resolution_2018, nspin_pos_2018, nspin_structure_2019, structure_2016, quantum_sensing_2017}.
One pre-requirement for such a sensing strategy is on-demand control of magnetic targets \cite{nmr_book_2013, nspin_structure_2019, 2D_nmr_2015}.
However, many phenomena in nature are neither magnetic nor controllable, and a new sensing paradigm needs to be developed.

\begin{figure*}
	\centering
	{\includegraphics[width=2.0\columnwidth]{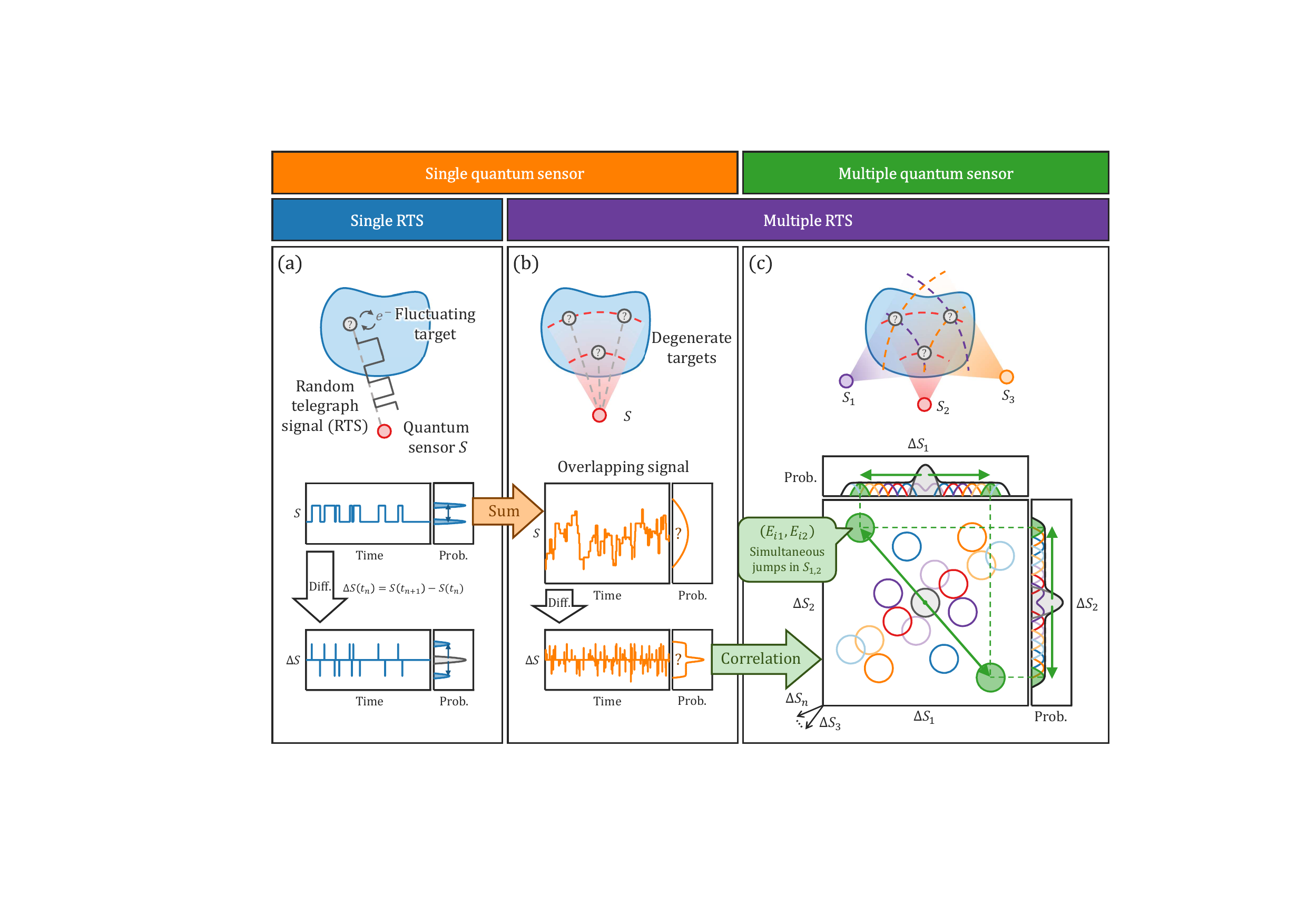}}
	\caption{The correlated sensing with multiple quantum sensors.
		(a) Schematic illustration of detecting a target with a random telegraph signal through a single quantum sensor. With a sensitive quantum sensor, the real-time dynamics and jump statistics are well detected.
		(b) The practical sensing situation with many targets simultaneously coupled to the single quantum sensor. The detected signal now is the summation of contributions from all the targets with a broad signal distribution, from which individual targets are not resolvable.
		(c) With spatially distributed multiple quantum sensors, each target induces random but simultaneous signal jumps for all the sensors. The signal amplitudes are position dependent and encode the structure information. The correlated detection of simultaneous jumps by multiple quantum sensors then determines a target with the corresponding amplitude vector ($E_{i1}$,$E_{i2}$,...,$E_{in}$).
		The circles displayed in the correlation spectrum denote the statistical distribution due to fluctuations or measurement errors in practical situations.
	}\label{fig1}
\end{figure*}

A general example with time-varying and uncontrollable signal is the random telegraph signal (RTS), which can be represented by many processes such as charge dynamics in semiconductors \cite{telegraph_2008, materials_challenges_2021}, transcriptional \cite{burst_2005, burst_2006, burst_2011} and electrophysiological activities in cells \cite{RTS_cancer_2020}, and protein conformational dynamics \cite{protein_dynamics_1991, protein_dynamics_2015, protein_dynamics_2018, protein_dynamics_2022}.
Fig.~\ref{fig1}(a) illustrates the simplest case: the quantum sensor is coupled to only one telegraph-like target. The dynamics of the target generate an RTS at the quantum sensor.
With sufficient sensitivity, the quantum sensor can resolve the RTS, and the amplitude of the RTS reveals the target's relative position to the quantum sensor.
However, in practical situations for complex structure analysis, many targets may simultaneously couple to the quantum sensor (Fig.~\ref{fig1}(b)). The overlapping of the signals makes individual targets indistinguishable.
Resolving the signals and extracting the underlying structural information remains challenging and is beyond the sensing capability of a single quantum sensor \cite{covariance_2022}.


Here, we propose a novel multi-sensor paradigm to resolve structural information that utilizes the signal correlation that is rarely exploited for quantum sensing.
In contrast to previous excellent works using the quantum entanglement among sensors for enhancing the sensing precision \cite{SQL_ion_2004, SQL_photon_2007, quantum_metrology_2011, SQL_atom_2016},
our method provides another avenue for boosting the structure reconstruction accuracy using signal correlation among sensors.
For spatially distributed quantum sensors (Fig.~\ref{fig1}(c)), each target induces correlated RTSs on all the quantum sensors with position-dependent amplitudes, respectively.
The correlation among signals reveals simultaneous jumps from a specific target as a clustering pattern in the high-dimensional differential signal space.
The resolving of clustering patterns relies on the drastic increase of the signal capacity, resembling the tomographic reconstruction: the clarity of the signal improves with an increase in the number of tomographic planes (the number of sensors here).
We experimentally illustrate the capability of correlated sensing with three NV centers based on their electrical detection.
By detecting the random telegraph electrical signals in a correlated way, we visualize the real-time random charge dynamics of 16 dark individual point defects and image them via a GPS-like localization method.
With these achievements, a full understanding of NV centers' optical spectral diffusion is obtained at the single charge level for the first time.

\begin{figure*}
	\centering
	{\includegraphics[width=2.0\columnwidth]{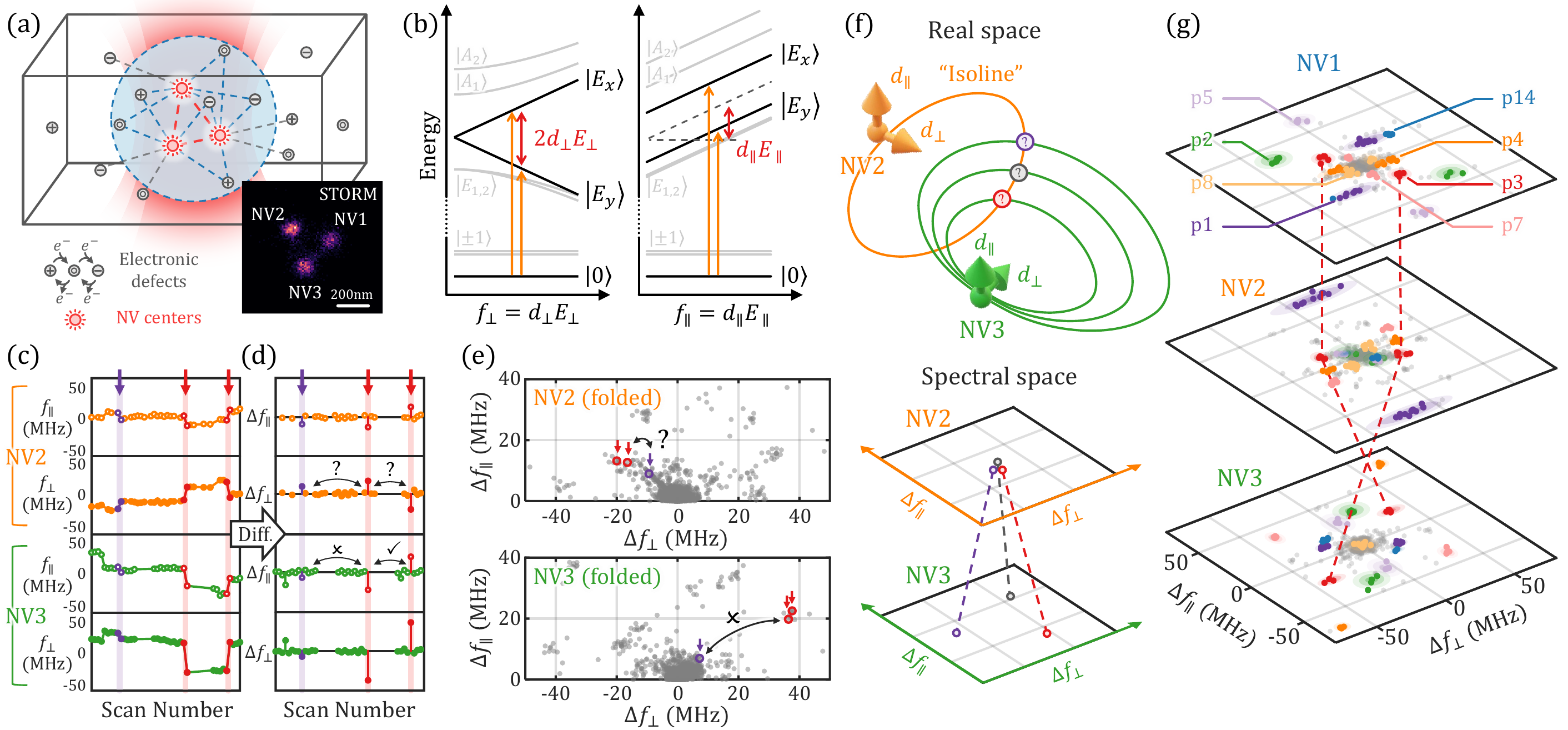}}
	\caption{Experimental system and correlated electrometry spectrum.
		(a) Schematic illustration of the correlated electrometry using multiple NV centers for electronic defect imaging. (Inset) A STORM super-resolution image of the three-NV system to verify the correlated electrometry.
		(b) The dependence of the excited $\left| E_x \right>$ and $\left| E_y \right>$ levels on the transverse ($E_\perp$) and longitudinal ($E_\parallel$) electric fields. $E_\perp$ component causes $\left| E_x \right>$ and $\left| E_y \right>$ to split, while $E_\parallel$ component causes $\left| E_x \right>$ and $\left| E_y \right>$ to shift simultaneously. The susceptibilities to the electric fields are $d_\parallel = 1.42$ MHz/(V/cm) and $d_\perp = 1.83$  MHz/(V/cm), respectively \cite{SM}.
		(c) Typical time-varying electric field signals from NV2 and NV3 obtained by consecutive PLE measurements. Three correlated jumps are labeled by arrows. Missing data points are due to the ionization of NV centers.
		(d) The differential of the signals in (c). From the NV2 jumps, the purple and red events have similar amplitudes (marked by the question marks), while they can be distinguished from the NV3 jumps (marked by the check mark and cross mark).
		(e) The joint distribution of all jumps of NV2 and NV3. ``Folded" indicates that the signs of the events with $\Delta f_\parallel<0$ are reversed due to the symmetry. The labeled jumps in (c,d) are highlighted. One can find the observed jumps with similar frequency shifts by NV2 (marked by the question mark) are separated in the NV3's spectral space (marked by the cross mark).
		(f) A geometric understanding of the correlated sensing to resolve point defects with two NV centers. In the 3D real space, all the point defects located on NV2's ``isoline" induce the same frequency shift and thus form an indistinguishable clustering pattern in the corresponding spectral space. But these locations are distinguished if NV3 is introduced. In NV3's spectral space, they map to distinct points.
		(g) A demonstration of 8 resolved point defects in the spectral spaces by considering all the correlations among the three NV centers. The two symmetric patterns for each defect correspond to charging and discharging events, respectively. The dashed lines indicate the spectral shifts from the same point defect.
	}\label{fig2}
\end{figure*}

\begin{figure*}
	\centering
	{\includegraphics[width=1.0\columnwidth]{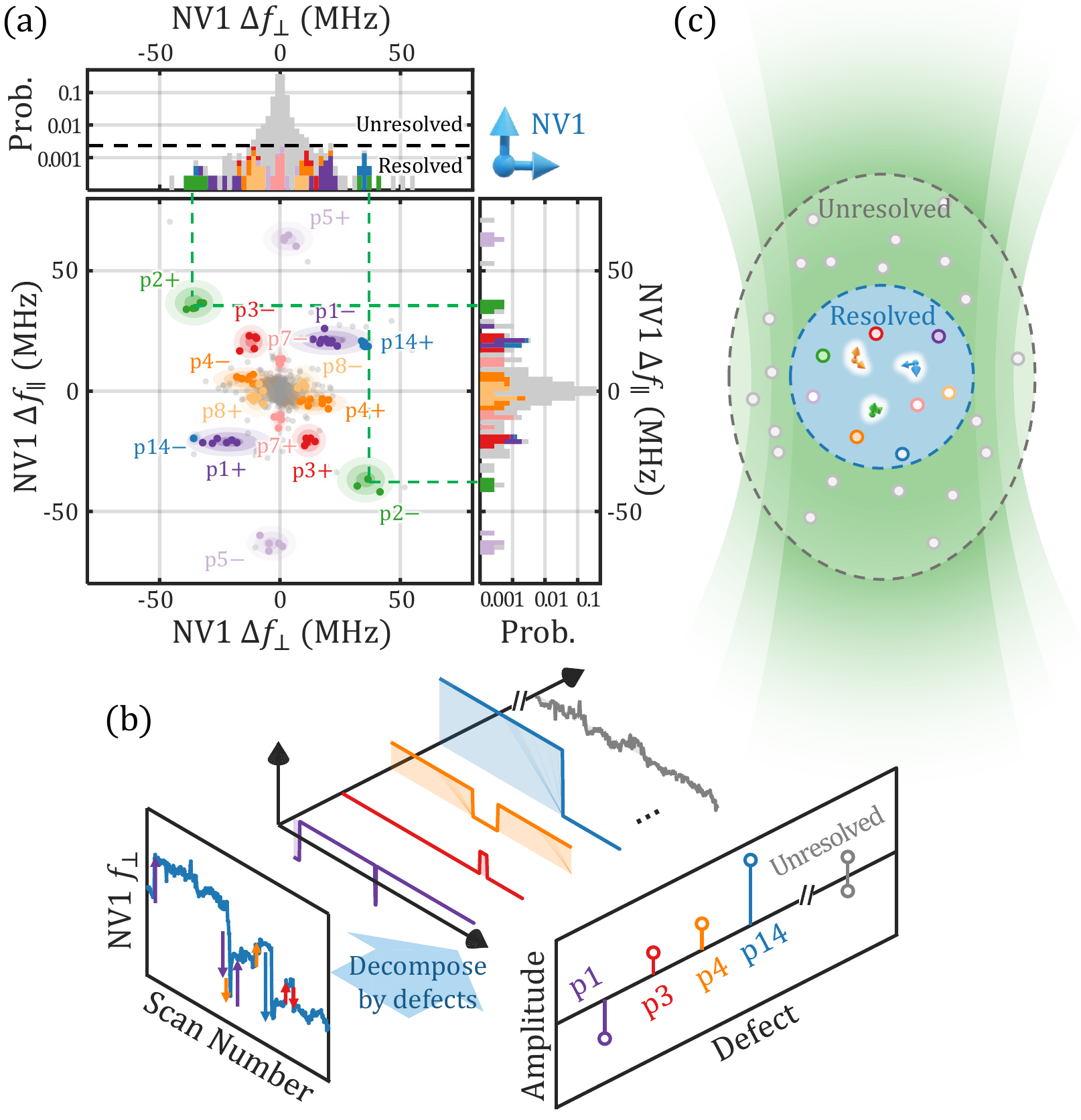}}
	\caption{Resolving individual defects and visualizing their real-time charge dynamics.
		(a) The spectral pattern of 8 resolved defects in NV1's spectral space (see Methods for the criteria for the resolved/unresolved cases). The resolved events are highlighted in the marginal distribution, and dashed lines mark the correspondence with the joint distribution. The defects are labeled with ``$+$" or ``$-$" signs, indicating the charging and discharging events.
		(b) The decomposition of NV1 $f_\perp$ signal by the resolved defects. (Left) The NV1 $f_\perp$ signal in the time domain. Arrows mark the major jumps caused by the resolved defects. (Top) The decomposed dynamics of the defects. The dynamics from distant defects are revealed with all the resolved defects' signals removed. (Right) The amplitudes of the signal from resolved and unresolved defects.
		(c) A geometric illustration of the resolved and unresolved defects.
	}\label{fig3}
\end{figure*}

The three NV centers used in our work are separated by approximately $200$ nm (Fig.~\ref{fig2}(a)), formed using the laser writing technique \cite{SM, laser_writing}, by which the position of NV centers is indeed highly controllable inside the diamond. Our sensing targets are the surrounding dark electronic defects, which have become one major noise source for various state-of-the-art solid-state quantum materials engineering \cite{materials_challenges_2021}.
Because of the finite band-gap of the substrate, the point defects inside usually experience undesired ionization and possess fluctuating charge states. These non-radiative pathways hinder their optical accessibility, rendering the detecting and studying of individual point defects challenging. On the other hand, the fluctuating charge dynamics represented as a random telegraph electrical signal opens the window for their electrical detection. In contrast to previous works relying on the ground spin state for electrical sensing \cite{electrical_sensing,electrical_sensing_du}, here we harness the excited state of NV centers as correlated electrometers to detect the electric field via the Stark effect \cite{electrical_sensing_2008, electrical_sensing_2011, electrical_sensing_2021} (Fig.~\ref{fig2}(b)) at the temperature of 11~K, which is about four orders of magnitudes more sensitive than its ground spin state \cite{electrical_sensing}.
Since its excited states $\left| E_x \right>$ and $\left| E_y \right>$ respond linearly to both the longitudinal and transverse electric field $E_\parallel$ and $E_\perp$, a single NV center acts as a 2D electric field sensor by performing the photoluminescence excitation spectroscopy (PLE) and fitting for transition frequencies.
Fig.~\ref{fig2}(c) displays a typical time-varying electric field signal obtained by consecutive PLE measurements of two NV centers. Both NV centers detect a series of simultaneous changes in electrical fields.
These correlated jumps with various amplitudes (Fig.~\ref{fig2}(d)) indicate that defects at different locations are either charging or discharging.
Defects in 3D space, however, are not always distinguishable by a 2D sensor alone, because all the defects on an ``isoline" are spectrally degenerate (Fig.~\ref{fig2}(f) top).
The further correlation between different sensors removes this degeneracy (Fig.~\ref{fig2}(f) down).
For example, the highlighted jumps clustered in NV2's spectral are distinguishable in NV3's spectral space (Fig.~\ref{fig2}(e)).
In other words, a defect repetitively induces spectral jumps and forms correlated clustering patterns in all the sensors' spectral space. With this criterion, from all the recorded electric field signals of all the three NVs, we can extract 16 defects' electric field amplitude at the three NVs' location \cite{SM}, and the spectral patterns of 8 defects are highlighted in Fig.~\ref{fig2}(g) for illustration.

The resolved defects' spectral patterns enable understanding the sensor's spectral jump statistics at the fundamental level and developing advanced control techniques.
The distribution of a single sensor's spectral jumps can be divided into two contributions: the resolved and unresolved fluctuations (Fig.~\ref{fig3}(a)).
The resolved fluctuations come from evident jumps that are caused by the charge dynamics of individual defects resolved by the sensors (Fig.~\ref{fig3}(b,c)).
The unsolved fluctuations are the residual fluctuations with much smaller amplitudes from distant defects.
Although these distant defects can not be spatially resolved, the induced fluctuations are still correlated.
A common-mode background electric field is reconstructed from the residual fluctuation signals from all three sensors \cite{SM}.
This strong correlation provides a new way to suppress spectral diffusion on demand.
For example, one can parallel monitor nearby NVs' spectrum and track or compensate for the electric field fluctuations without interrogating and disturbing the operating NV.

\begin{figure*}
	\centering
	{\includegraphics[width=2.0\columnwidth]{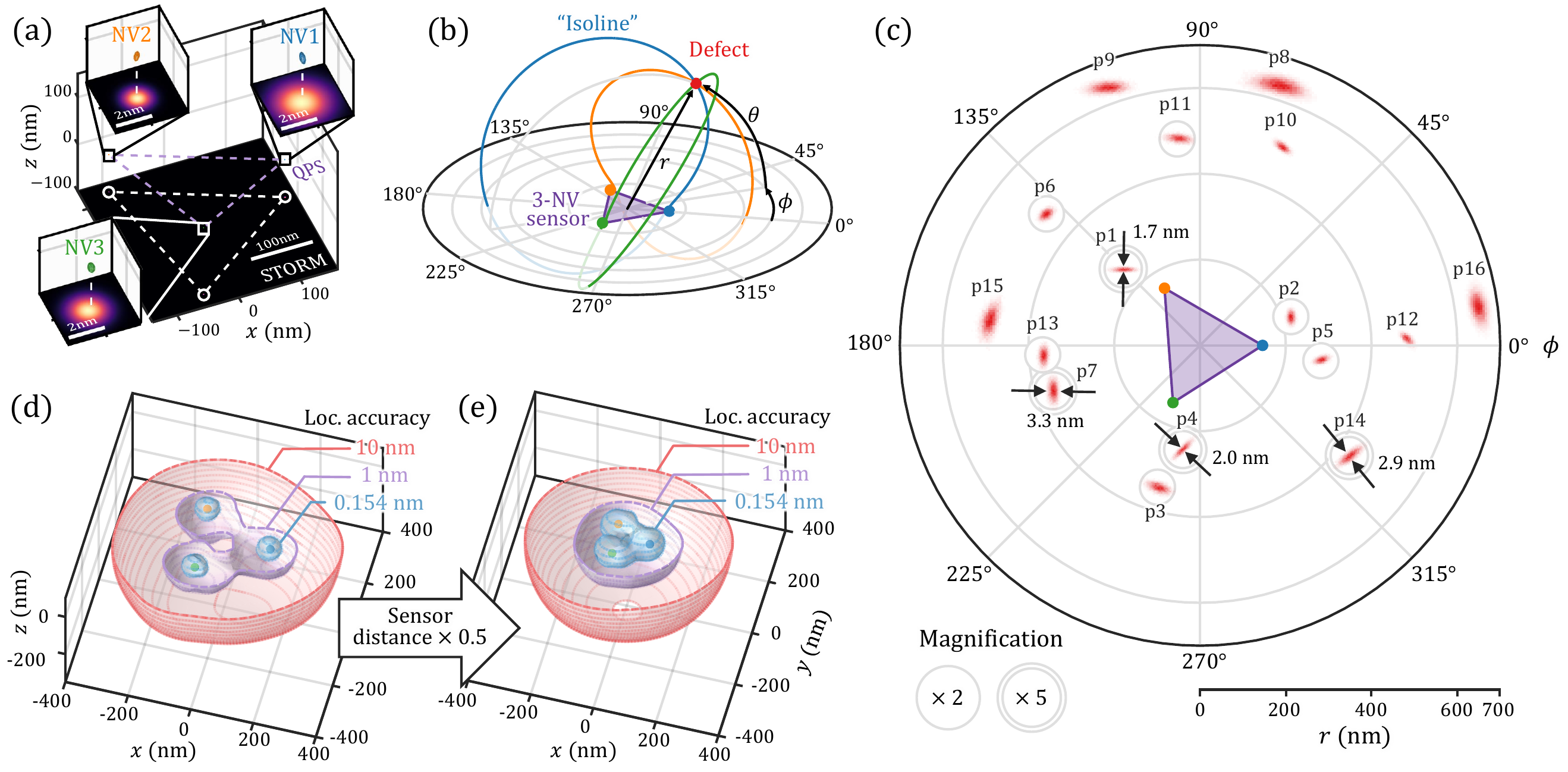}}
	\caption{Localization with the quantum positioning system (QPS).
		(a) The comparison between the NV centers' STORM and QPS localization results. The white (purple) dashed lines mark the STORM (QPS) results. Sample rotation and tilts between different setups are taken into account. Insets are zoom-in views at the NVs. The ellipsoids correspond to the FWHM of the probability distribution of the NV positions. The STORM localization results of the NVs are displayed in the $xy$-plane.
		(b) A geometric illustration of the QPS localization method. The position of the defect is determined by the three isolines corresponding to the electric field amplitudes at the three NVs.
		(c) An image of the resolved defects. Only the spherical coordinates $r$ and $\phi$ of the defects are shown. The density indicates the probability distribution of the localized defect. Some defects are magnified for clarity. The FWHM of p1, p4, p7, and p14 along their best-localized directions are denoted and marked by arrows.
		(d,e) The estimation of localization accuracy. The outer, middle, and inner surfaces correspond to the isosurfaces with a localization FWHM of 10 nm, 1 nm, and 0.154 nm, respectively, caused by a spectral measurement error of 1.0 MHz \cite{SM}. (d) corresponds to the three-NV system in our work, and (e) corresponds to a three-NV system with half the distance.
	}\label{fig4}
\end{figure*}

Finally, we develop a GPS-like localization method, namely the quantum positioning system (QPS), to spatially 3D image the resolved defects (not to be confused with quantum-enhanced positioning system \cite{QPS_2001}).
QPS localizes individual defects using their electric fields (Fig.~\ref{fig4}(b)), similar to satellite navigation systems using distance to determine the target location.
With three NV centers, the position of the defect is located at the intersection point of the three corresponding isolines (Fig.~\ref{fig2}(f)) with high accuracy.
This method is first verified by determining the relative location between NV centers and observing the consistency with the STORM super-resolution image (Fig.~\ref{fig4}(a)).

With our QPS system, the 16 detected defects distributed in a micron-scale region are located and displayed in Fig.~\ref{fig4}(c).
The defect with a distance of 125 nm (from the nearest NV) is localized with a full-width at half-maximum (FWHM) of 1.7 nm, generally increasing with the distance increase. The uncertainty is mainly due to statistical errors.
To estimate our method's localization accuracy, we consider the experimental spectral measurement error of 1.0 MHz \cite{SM}. We define the average standard error as $\overline\sigma_{3d}=\sqrt[3]{\sigma_1 \sigma_2 \sigma_3}$, and the localization accuracy as the FWHM $ 2\sqrt{2 \ln 2} \overline\sigma_{3d}$. Here $\sigma_{1,2,3}$ are the standard error along the three principal axes of the 3D error ellipsoid \cite{SM}.
The detection regions of the three-NV QPS with localization accuracy better than 10 nm, 1 nm, and 0.154 nm (i.e., the length of the diamond C-C bond) are shown in Fig.~\ref{fig4}(d).
To further increase the localization accuracy in the proximity of the three-NV system (i.e., the 0.154 nm region), QPS with denser sensors can be designed to shorten the distance from the NVs to the target defect.
As shown in Fig.~\ref{fig4}(e), the detection region with the localization accuracy of 0.154 nm will expand without losing much of the 10 nm region's detection range, for QPS with half the NVs' distances of those in Fig.~\ref{fig4}(d).

In conclusion, we develop a correlated sensing paradigm with a quantum multi-sensor system for atomic-scale structural analysis. This method fully explores the signal correlation among sensors and suits for detecting another type of random signal that widely exists in nature. The capability of this method is demonstrated by performing correlated optical electrometry with three NV centers to obtain nanoscale imaging and real-time charge dynamics of the electronic point defects in diamonds. Two immediate applications for our method can be expected.
First, the further advancement of current solid-state quantum materials enters the stage of point defect engineering with a density much below the state-of-the-art detection techniques \cite{materials_challenges_2021}. Our results demonstrate that a defect density of 0.01 ppb level is detectable, which is about two orders of magnitude better than secondary ion mass spectrometry, currently the most sensitive method for heteroatomic detection. The highly sensitive detection of defects is crucial not only for advancing quantum technologies \cite{quantum_defect_2021}, but also for semiconductor devices, of which defects are inevitably introduced in the manufacturing processes. 
The only requirements, the optical emitters and the susceptibility to the electric field, mean that our method can be directly applied to other solid-state materials, like silicon-carbide and silicon.
Second, our method extends the sensing capacity by correlating multiple sensors and harnessing the random nature of the targets. Our method provides a new way of distinguishing among numerous fluctuating targets. The drastic increase in the sensing capacity complies with the increasing target number for complex structural analysis. 



%

\textbf{Acknowledgments.} This work is supported by the National Natural Science Foundation of China (Grants No.\ 92265204, T2325023, T2125011, 12104446, 12104447), the CAS (GJJSTD20200001), the Innovation Program for Quantum Science and Technology (Grant No. 2021ZD0302200), the Anhui Initiative in Quantum Information Technologies (Grant No. AHY050000), the Fundamental Research Funds for the Central Universities.

\textbf{Author contributions:} J.D. and Y.W. supervised the project. Y.W. designed the experiments. W.J., Z.L., Y.G. performed the experiments. J.Z., P.Y., M.W prepared the sample. Z.H., Y.G., K.X. performed the STORM measurements. W.J., Z.L., Z.H., S.D., Y.C. carried out the calculations. Y.W., W.J., J.D., F.S. wrote the manuscript. All authors discussed the results and commented on the manuscript.

\textbf{Competing interests:} The authors declare no competing interests.

\textbf{Data and code availability:} All data and code that support the findings of this study are available from the corresponding author upon reasonable request.

\section*{Methods}

\subsection{I. Sample information}

The sample used in this work is fabricated from an ``electronic grade'' single-crystal diamond plate from Element Six Ltd., with a nitrogen concentration of $<$5ppb. The sample is cut such that the top surface is (001) crystal plane and the side surfaces are in \{110\} crystal plane family. The NV centers are generated by a home-built femtosecond laser writing setup, in which short high-power laser pulses are focused in the sample to create vacancies in the diamond lattice. A train of 808 nm fs pulses is generated by a Ti:Sapphire laser (Coherent Chameleon Ultra II) and chopped by a pulse picker (Eksma UP2) down to single pulses. To optimize the diamond lattice's damage, the laser pulse's wavefront distortion is compensated by pre-distorting using a deformable mirror (Alpao DM97-15) combined with a 4f imaging configuration. After writing the vacancies, the sample is annealed for 4 hours at 1300K to form NV centers in the writing spots. The three-NV system is 4.7 $\mu$m below the surface. Note that with this highly controllable femtosecond laser writing technique, multi-sensor systems can be reliably reproduced, and systems with more and denser sensors can be readily engineered.

\subsection{II. Experimental setup}

The QPS experiments in this work are performed on a home-built low-temperature confocal setup. The sample is hosted at the temperature of 11 K in a closed-cycle optical cryostat (Montana Instruments Cryostation S200). The sample is positioned by an XYZ piezo positioner and scanner stack (Attocube) at the focal point of a 0.9 NA objective (Olympus MPLFLN100x), of which the temperature is stabilized to 300 K. We use three lasers to excite the NV center: a 532 nm laser (Changchun New Industries Optoelectronics Technology) to reset or disturb the charge state of NVs and defects, and two tunable 637 nm lasers (New focus TLB-6704-OI, Toptica DLC DL PRO HP 637) to perform resonant excitation. The wavelengths of the 637 nm lasers are stabilized using a wavemeter (HighFiness WSU-10). 
Microwave pulses are generated by IQ mixing the signal generated by an arbitrary waveform generator (Zurich Instruments HDAWG-8) and a vector signal generator (Ceyear Technologies 1435B-V Vector Signal Generator). After amplification (minicircuit), the microwave pulses are fed to a gold strip line fabricated on top of the sample. An arbitrary sequence generator (CIQTEK ASG8005) controls the timing of the laser and microwave pulses.

The STORM measurements are performed on a home-built confocal-STORM setup. The confocal imaging is achieved by scanning a 1.45 NA objective lens (OLYMPUS UPlanXApo100x) using a piezo scanner (Coremorrow N12.XYZ100S-D2) under the illumination of a 532 nm laser and collecting fluorescence using an APD (Excelitas Technologies SPCM-AQRH-44-BR1). The STORM imaging is performed by illuminating the sample with a 594 nm laser (Changchun New Industries Optoelectronics Technology) and capturing the fluorescence image continuously with an sCMOS (Dhyana 400BSI).

\subsection{III. NV centers as electric field sensors}

The three-NV QPS is constructed by three NV centers, each of which is a 2-dimensional electric field sensor.
The effective Hamiltonian for the $\{\ket{ E_x }, \ket{ E_y }\}$ subspace of the excited state manifold is given by the effective Hamiltonian
\begin{equation}
	H_\mathrm{eff} = d_\parallel E_z + d_\perp \left(
	\begin{array}{cc}
		E_x & -E_y \\
		-E_y &  - E_x
	\end{array} \right),
\end{equation}
in which $d_{\parallel,\perp}$ are the susceptibilities to the electric field $E_{x,y,z}$.
As shown in Fig.~\ref{fig1}(f), states' energy responds linearly to the electric field.
With the transition frequencies of $\ket{0} \leftrightarrow \ket{E_x}$ and  $\ket{0} \leftrightarrow \ket{E_y}$ measured, namely $f_{\ket{E_x}}$ and $f_{\ket{E_y}}$, the electric field can be obtained by
\begin{equation}
	\begin{aligned}
		f_\parallel &= \frac{f_{\ket{E_x}} + f_{\ket{E_y}}}{2} = d_\parallel E_z ,\\
		f_\perp &= \frac{f_{\ket{E_x}} - f_{\ket{E_y}}}{2} = d_\perp |\boldvec{E}_\perp| .
	\end{aligned}
\end{equation}
For a small perturbation field $\Delta {\boldvec{E}}$ compared to the static strain field ${\boldvec{E}}_0$, spectroscopy measurements respond approximately linearly to two components of $\Delta {\boldvec{E}}$: i) the longitudinal component along NV axis direction $\hat{n}_\parallel$ as
\begin{equation}
	\Delta f_\parallel = d_\parallel \Delta E_z = d_\parallel \Delta \boldvec{E} \cdot \hat{n}_\parallel,\\
\end{equation}
and ii) the transverse component along the transverse field $\boldvec{E}_{0,\perp}$ direction $\hat{n}_\perp$ as
\begin{equation}
	\Delta f_\perp = d_\perp \left( |\boldvec{E}_{0,\perp} + \Delta \boldvec{E}_\perp| - |\boldvec{E}_{0,\perp}|\right) \approx d_\perp \Delta \boldvec{E} \cdot \hat{n}_\perp,
\end{equation}
making the NV center a 2-dimensional sensor to the perturbation field $\Delta {\boldvec{E}}$.

The perturbation field can come from the charge fluctuations of NVs and defects, while for clarity, all the spectral results shown in the main text are with the NVs' electric fields removed.
To discriminate spectral jumps originating from the same defect, there are three criteria: (i) the jump events occur repetitively, (ii) the spectral jumps occurring at the three NVs are correlated, and (iii) the localization result passes the hypothesis test given in Supplementary Material Sec. II \cite{SM}.
The original spectral results and the extracted electric field components of the NVs' charges and all the resolved defects are given in Supplementary Material Sec. III \cite{SM}.

\subsection{IV. QPS localization}

To determine the location of a defect, we use the longitudinal and transverse spectral components from the defect to NV$i$ as observables, i.e.,
\begin{gather}
	F_{i,\parallel}\left(\boldvec{r}\right)
	= -\frac{d_{\parallel}\hat{\boldvec{z}}_i \cdot \left(\boldvec{r} - \boldvec{r}_i\right)} { 4\pi\epsilon \left|\boldvec{r} - \boldvec{r}_i\right|^3}, \\
	\begin{align}
		& F_{i,\perp}\left(\boldvec{r}\right) = \\
		& \sqrt{ \sum_{\hat{\boldvec{n}}=\{\hat{\boldvec{x}}_i,\hat{\boldvec{y}}_i\}}
			\left(d_{\perp}\boldvec{E}_{i, \perp} \cdot \hat{\boldvec{n}} - \frac{d_{\perp} \hat{\boldvec{n}} \cdot \left(\boldvec{r}-\boldvec{r}_i\right)} { 4\pi\epsilon \left| \boldvec{r} - \boldvec{r}_i \right|^3} \right)^2}
		- |d_{\perp} \boldvec{E}_{i, \perp}|, \nonumber
	\end{align}
\end{gather}
in which $\hat{\boldvec{x}}_i$, $\hat{\boldvec{y}}_i$ and $\hat{\boldvec{z}}_i$ are the unit vectors of local coordinate of NV$i$, $\boldvec{r}$ and $\boldvec{r}_i$ correspond to the position of the target defect and the NV$i$, and $\epsilon$ is the dielectric constant of diamond.
The problem of localizing a defect with the three-NV QPS is overdetermined.
The location of the defect is determined by minimizing the sum of the squared residuals in the observables, as given by the objective function
\begin{align}
	& O\left(\boldvec{r}\right) = \\
	& \sum _{i=1}^{n} \left[ \left(\frac{F_{i,\parallel}\left(\boldvec{r}\right)-\Delta f_{i,\parallel}}{\sigma_{f_{i,\parallel}}}\right)^2
	+ \left(\frac{F_{i,\perp}\left(\boldvec{r}\right)-\Delta f_{i,\perp}}{\sigma_{f_{i,\perp}}}\right)^2 \right], \nonumber
\end{align}
in which $\sigma_{f_{i,\parallel}}$ and $\sigma_{f_{i,\perp}}$ correspond to the measurement errors of the longitudinal and transverse spectral components.
A general discussion of the localization problem with multiple NVs is given in Supplementary Material Sec. II \cite{SM}, and the localization results of all the experimentally resolved defects with the three-NV QPS are given in Supplementary Material Sec. V \cite{SM}.

\begin{figure*}
	\centering
	\includegraphics[width=1.85\columnwidth]{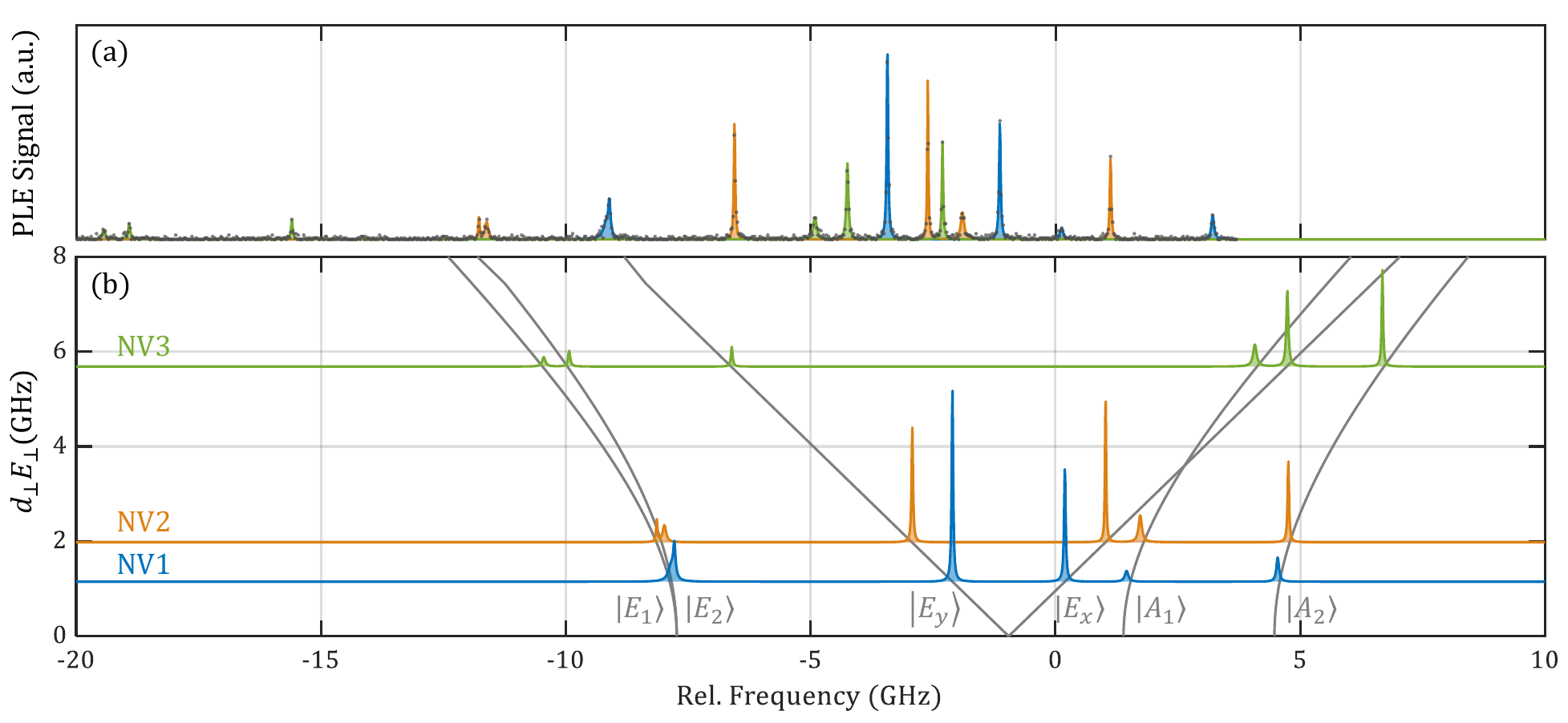}
	\caption{(Extended data)
		PLE spectra fit results for the three NVs. 
		(a)~The full spectrum of the three NVs.
		(b)~Fit the spectra for the transverse strain field components. The longitudinal components are set to zero.
	}\label{fig: PLE fit}
\end{figure*}

\begin{figure*}
	\centering
	\includegraphics[width=2\columnwidth]{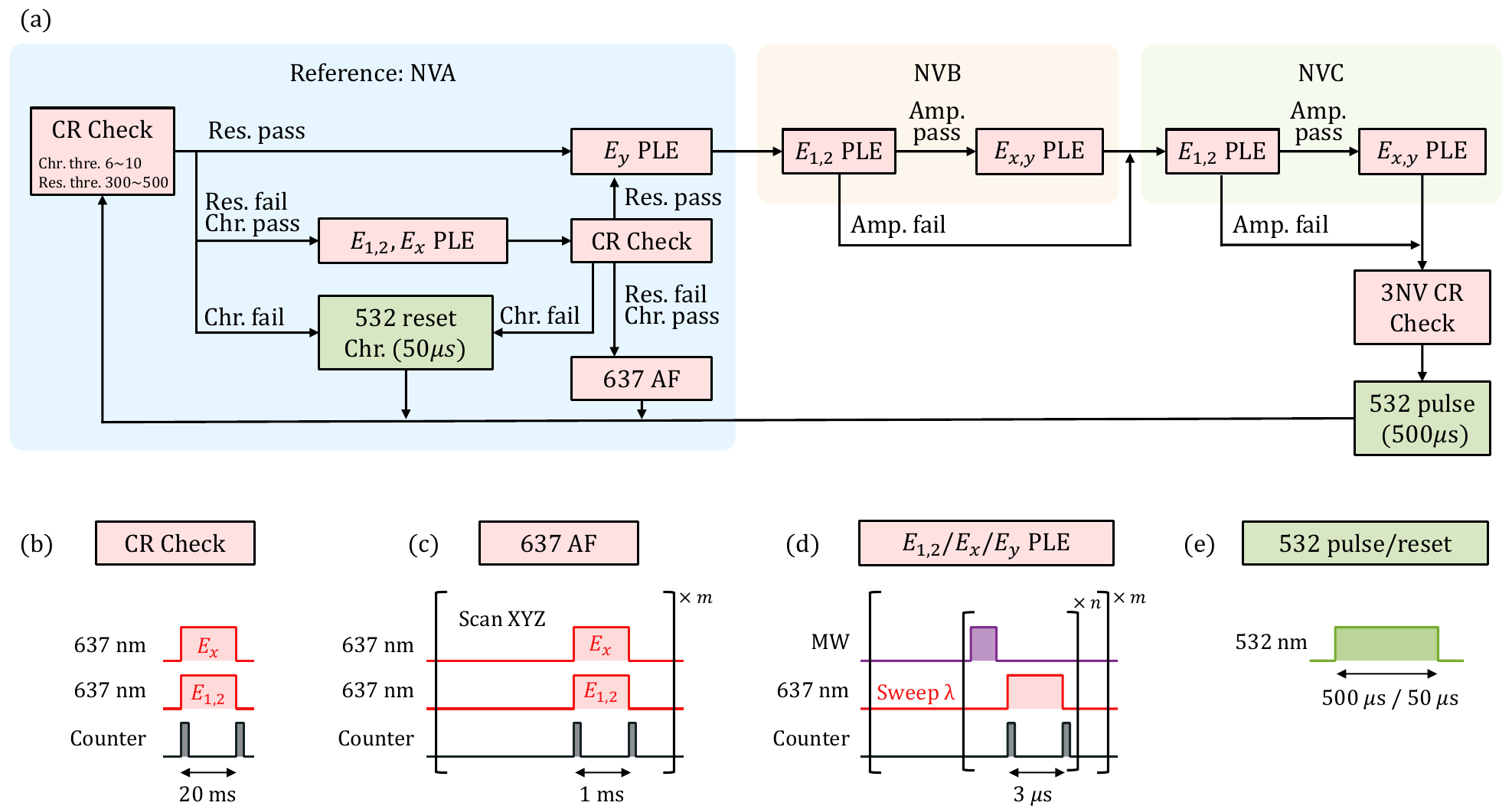}
	\caption{(Extended data)
		The flowchart and pulse sequences for the experiments.
		(a)~The flowchart of the experiments control logic. 
		One of the three NVs is chosen as a reference and labeled as NVA; the other two NVs are labeled as NVB and NVC. 
		A single experiment run is started by checking the resonance and charge state of NVA, namely ``CR check" \cite{entanglement_2013}. 
		If NVA passes the resonance check, then the PLE of NVA $E_y$ is measured. 
		If NVA fails the charge check, a 50 $\mu$s 532 nm laser pulse is applied to reset the charge state of NVA. 
		If NVA fails the resonance check but passes the charge check, then $E_x$ and $E_{1,2}$ PLE spectra are measured to calibrate the transition wavelengths. 
		After the wavelength calibration, if NVA still fails resonance check, an auto-focusing process with $E_x$ and $E_{1,2}$ lasers is applied, labeled as ``637 AF".
		These steps are repeated until NVA passes the resonance check, ensuring NVA's resonance and position are tracked in the experiments.
		After measuring NVA's $E_y$ PLE spectrum, the spectra of NVB and NVC are measured.
		For NVB and NVC, $E_{1,2}$ PLE spectrum is first measured to determine the charge state.
		If the $E_{1,2}$ peak is missing, the subsequent $E_x$ and $E_y$ measurements will be skipped.
		After measuring all the spectra, another round of CR checks for the three NVs is applied to detect ionization events during the PLE measurements. 
		Finally, a 532 nm laser pulse of 500 $\mu$s is applied to disturb the defect charge state deliberately. 
		(b-e)~The pulse sequences for the steps in (a). 
	}\label{fig: flowchart}
\end{figure*}

\begin{figure*}
	\centering
	\includegraphics[width=2.0\columnwidth]{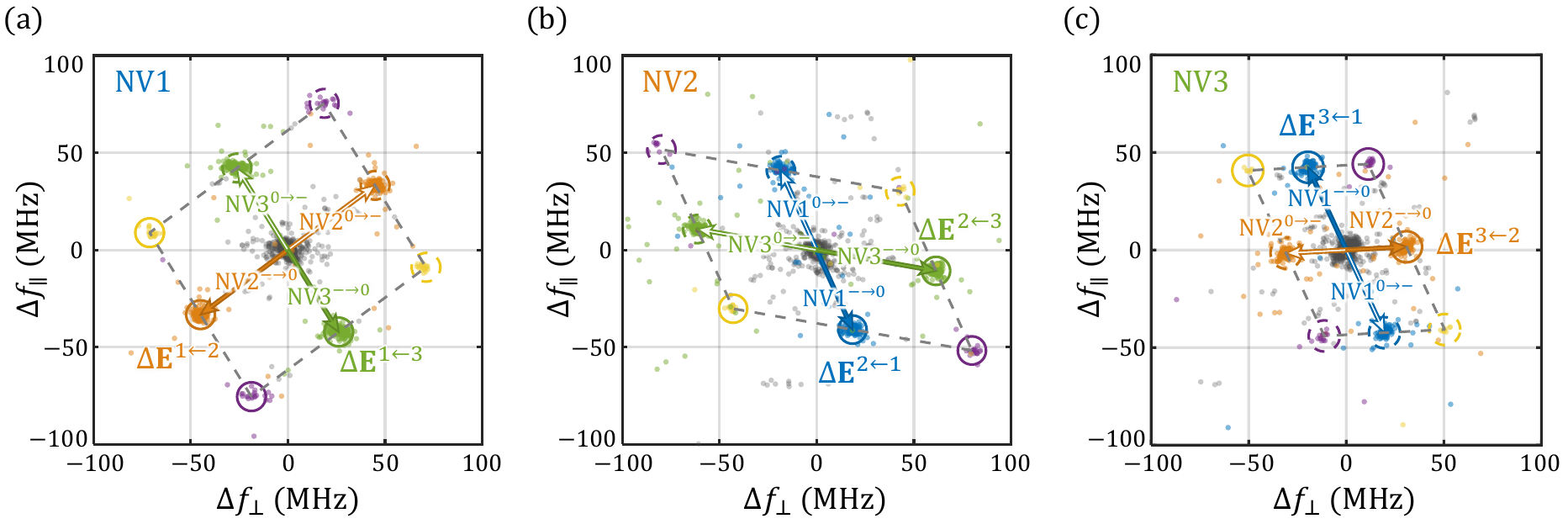}
	\caption{(Extended data)
		The 2D differential spectra of (a) NV1, (b) NV2 and (c) NV3. The color of the points indicates the charge state change of the NVs. The most evident pattern here is caused by the electric field from the other two NVs, as labeled by the arrows.
		To reveal the defects' pattern, the spectra results shown in the main text are with the NVs' electric field removed.
	}\label{fig: NV dE}
\end{figure*}

\begin{figure*}
	\centering
	\includegraphics[width=2\columnwidth]{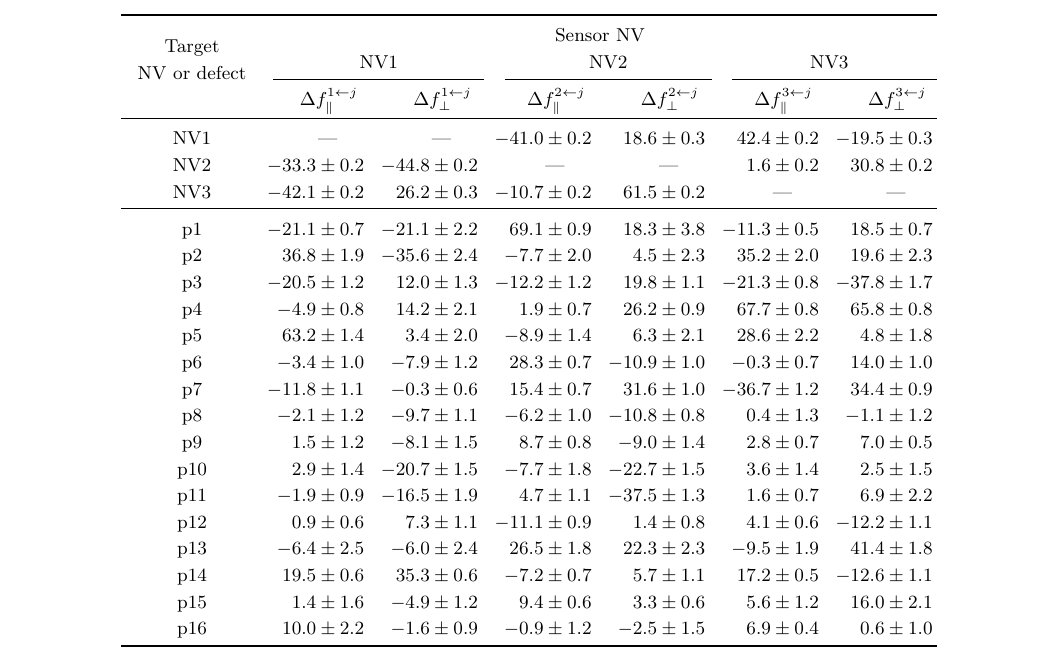}
	\caption{(Extended data)
		A summary of differential spectroscopy results of NVs and defects. 
		$\Delta f_{\parallel,\perp}^{i \leftarrow j}$ denotes the spectral shift of NV$i$ caused by adding a charge at target $j$.
		The unit of the data is MHz, and the errors correspond to the 95\% confidence interval.
	}\label{tab: diff. spec. charge}
\end{figure*}

\begin{figure*}
	\centering
	\includegraphics[width=1.5\columnwidth]{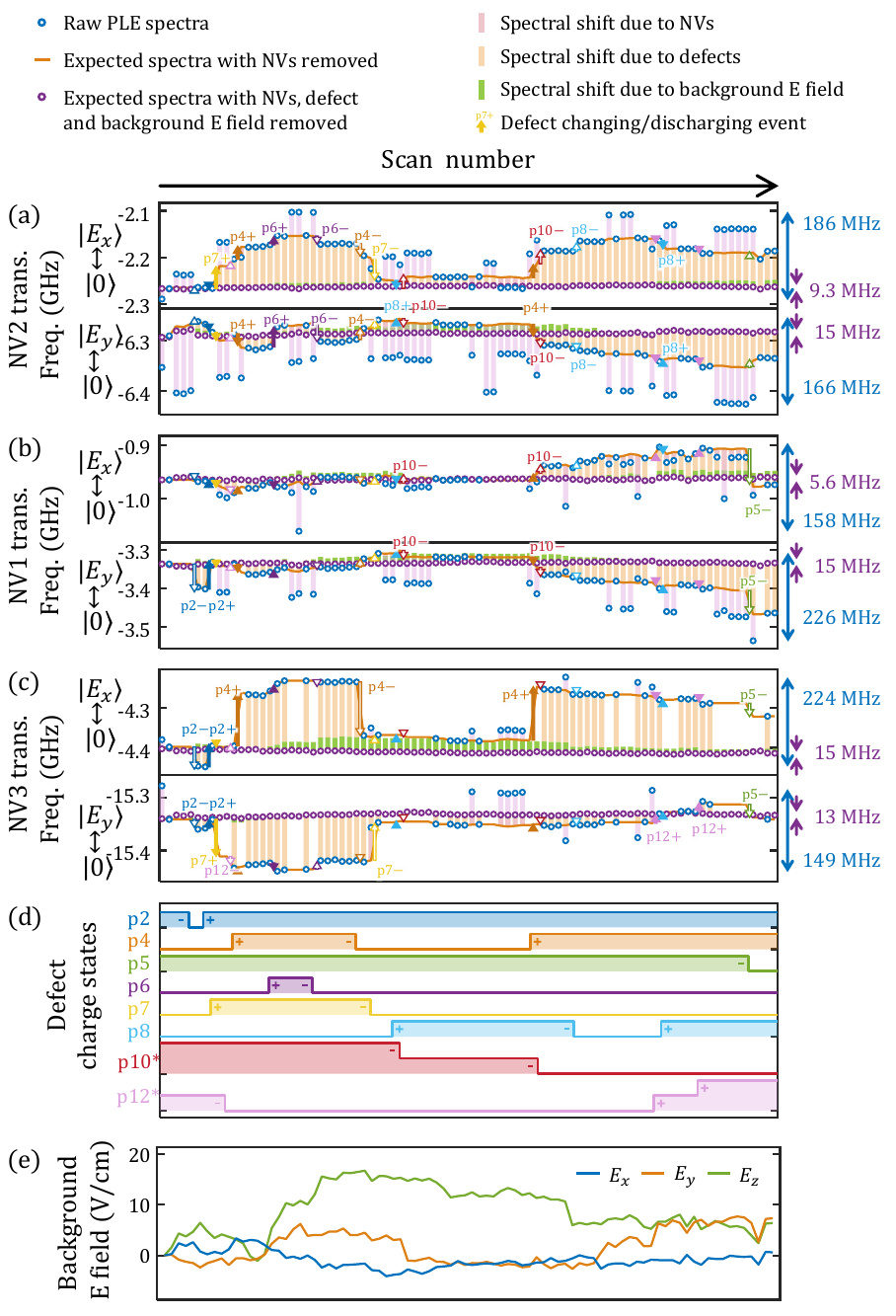}
	\caption{(Extended data)
		The decomposition of the raw PLE spectra of the three NVs.
		(a-c) The raw PLE spectra within consecutive scans of the $E_x$ and $E_y$ transitions of NV2, NV1 and NV3 (in the experiment order).
		The blue data points are the raw experimental results of PLE spectra peak position with evident spectral diffusion of a peak-to-peak variation around 200 MHz.
		The pink, orange and green bars indicate the contributions of spectral shifts due to NVs' charges, resolved defects' dynamics (d) and background electric field from distant defects (e).
		The purple data points are the expected spectra with all the shifts corrected, which are nearly diffusion-free with a peak-to-peak variant close to the natural linewidth of 13 MHz.
		Arrows mark the spectral shifts caused by the resolved defects.
		(d) The resolved defect charge dynamics. The p10* and p12* are possibly defects with at least three charge states or two indistinguishable charge defects.
		(e) The common-mode background electric field reconstructed by the three NVs.
	}\label{fig: decompose}
\end{figure*}

\begin{figure*}
	\centering
	\includegraphics[width=2\columnwidth]{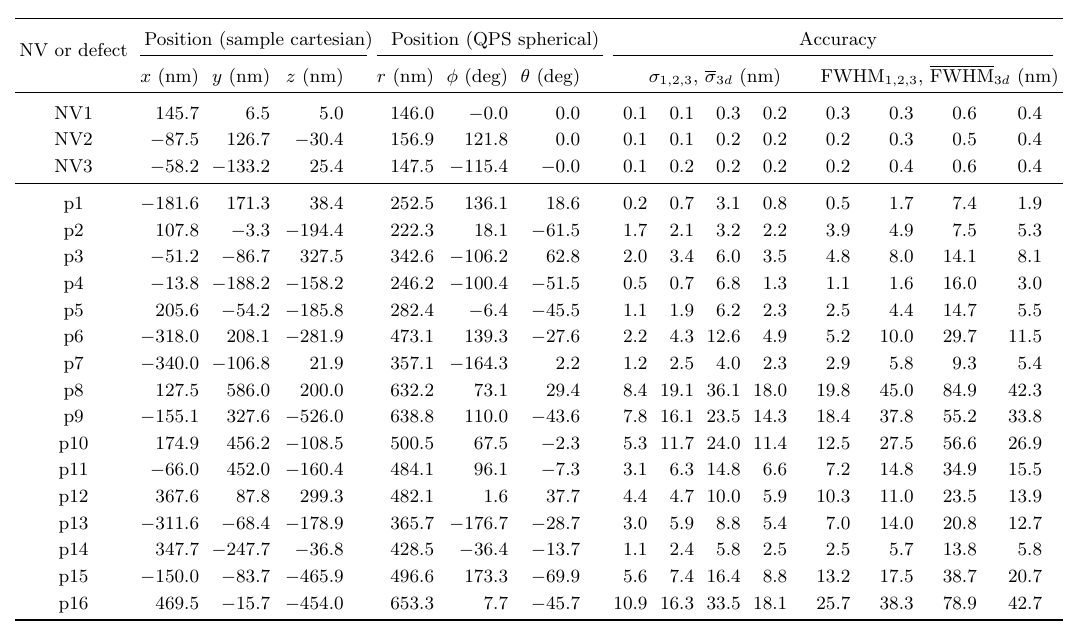}
	\caption{(Extended data)
		A summary of the localization results of all the NVs and defects. The positions are given in two coordinate systems: the sample cartesian coordinate and the QPS spherical coordinate (defined by the three-NV system). The localization accuracies are presented in standard errors along the principal axes $\sigma_{1,2,3}$, the geometric average $\overline\sigma_{3d}$, and their corresponding FWHM.
		Note that in the best-localized direction of p1, the FWHM is 0.5~nm, better than the 1.7~nm demonstrated in Fig.~\ref{fig4}(c). This is due to projecting to the $(r,\phi)$-plane in Fig.~\ref{fig4}(c).
	}\label{tab: charge loc.}
\end{figure*}

\end{document}